\documentclass[12pt,preprint]{aastex}

\slugcomment{To appear in ApJ (2002 July 20)}

\shorttitle{H I in the Helix}
\shortauthors{Rodr\'\i guez, Goss, \& Williams}

\begin{document}


\title{VLA Observations of H I in the Helix Nebula (NGC~7293)}


\author{Luis F. Rodr\'\i guez}
\affil{Instituto de Astronom\'\i a, Campus UNAM,
Morelia, Michoac\'an 58190, M\'exico}
\email{l.rodriguez@astrosmo.unam.mx}

\author{W. M. Goss}
%\altaffilmark{1}} 
\affil{National Radio Astronomy Observatory, 
P.O. Box O, Socorro, NM 87801-0387}
\email{mgoss@nrao.edu}

\and

\author{Robert Williams}
\affil{Space Telescope Science Institute, 3700 San Martin Drive, 
Baltimore, MD 21218}
\email{wms@stsci.edu}


%\altaffiltext{1}{Visiting Astronomer, Cerro Tololo Inter-American Observatory.
%CTIO is operated by AURA, Inc.\ under contract to the National Science


\begin{abstract}
We report the detection of 21-cm line emission from H~I in the
planetary nebula NGC~7293 (the Helix). The observations, 
made with the Very Large Array, show the presence of a ring of atomic hydrogen
that is associated with the outer portion of the ionized nebula. 
This ring is most probably gas ejected in the AGB phase that
has been subsequently photodissociated by radiation from the central star.
The H~I emission spreads over $\sim$50 km s$^{-1}$ in radial
velocity.
The mass in H~I is $\sim$0.07 M$_\odot$, about three times larger than
the mass in molecular hydrogen and comparable with the mass in
ionized hydrogen.
\end{abstract}




\section{Introduction}

Planetary nebulae (PNe) are known to contain significant amounts
of molecular and neutral atomic gas, in addition to the
ionized component that characterizes them.
Carbon monoxide was first detected in NGC~7027 by Mufson, Lyon, \&
Marionni (1975),
while atomic hydrogen was first detected in NGC~6302
(Rodr\'\i guez \& Moran 1982). At present there are about
60 PNe detected in the millimeter lines of CO (Huggins et al. 1996)
or in near-infrared H$_2$ emission (Kastner et al. 1996). 
However, only about one dozen PNe have been detected in the 21-cm line
(Gussie \& Taylor 1995; Rodr\'\i guez, G\'omez, \& L\'opez 2000).
It is unclear if these non detections
imply that atomic hydrogen is less common
than molecular hydrogen in PNe, but most likely this is
just the result of contamination from line-of-sight galactic H~I,
confusing the detection of H~I in the PNe.
Indeed, the PNe detected in H I are usually at relatively high galactic
latitudes or have peculiar radial velocities that allow
the observer to disentangle the emission associated
with the planetary nebula from that
of line-of-sight H~I features.

The Helix Nebula (NGC~7293, PK~36-57.1) is one of the best studied
PNe, with its bright 
emission from ionized gas
having an angular diameter of $\sim$15$'$, that at a distance
of 200 pc (Harris et al. 1997) implies a radius of $\sim$0.4 pc.
Several observations (i. e. O'Dell 1998; Speck et al. 2002)
show faint emission extending over twice the angular size
of the classical bright nebula. 
The large angular size makes it an ideal object for the study
of the spatial distribution and stratification 
of the different components (ions, atoms, molecules, and dust)
that may be present in and surrounding the PN.

O'Dell (1998) and Henry, Kwitter, \& Dufour (1999) propose that
the Helix nebula appears to be a thick disk of material with its
plane nearly face-on.
We will adopt this geometry in our discussion.
In their recent study of CO emission from the Helix,
Young et al. (1999) conclude that
the ionized gas abuts the molecular gas, indicating strong stratification
effects in the spatial distribution of the neutral and ionized components.

From the above results, O'Dell, Henney, \& Burkert (2000) conclude that
the Helix is optically thick to ionizing photons (i. e. ionization
bound) in the plane of
this disk, whereas it is probably optically-thin to ionizing photons 
(i. e. density bound) in directions close to its polar axis.
In contrast, from observations of molecular
hydrogen, Speck et al. (2002) show that the molecular species
coexist with the ionized gas on a scale of 0.1-0.3 pc,
and note that this requires that
the basic structure of the nebula must be due to the strong
stratification that occurs within individual, small-scale ($\sim 1''$)
inhomogeneities.

In this paper we present VLA observations of the Helix Nebula
that indicate associated H I 
and allow the determination of the spatial distribution and
of the total mass in the H~I component.

\section{Observations}

The observations were made on 1995 January 25
using the Very Large Array (VLA) of
the National Radio Astronomy Observatory (NRAO)\footnote{NRAO is
a facility of the National Science Foundation
operated under cooperative
agreement by Associated Universities, Inc.}\ in the DnC
configuration.
One IF was used for the continuum observations, with a
bandwidth of 12.5 MHz, two polarizations, and centered at 1435.2 MHz.
The other IF, also with two polarization,
was used for the line observations, centered
at the rest frequency of the hyperfine transition of H I
(1420.406 MHz), with a total of 127 channels over a total bandwidth
of 0.78 MHz (1.29 km s$^{-1}$ per channel). After
Hanning weighting, this resulted in a velocity resolution
of 2.58 km s$^{-1}$ per channel. The observations were centered at 
$v_{LSR} = -$24 km s$^{-1}$, close to the systemic velocity of the 
CO emission from the planetary nebula, $v_{LSR} = -$23 km s$^{-1}$  
(Young et al. 1999)
The data were edited, calibrated, and imaged using the software
package Astronomical Image Processing System (AIPS) of NRAO.
 
\subsection{Continuum Observations}

We show in Figure 1 the continuum image of the Helix at 1.4 GHz
from the 12.5 MHz IF.
The synthesized beam has dimensions of
54$\rlap.{''}2$$\times$39$\rlap.{''}3$; PA=
-5$\rlap.^\circ$0.
The familiar ring morphology of the PN is easily recognized.
Several bright compact radio sources are also evident in the image.
They are most probably background non-thermal objects.
The total flux density of the Helix (obtained integrating over the face of the
nebula in an image corrected for the
primary beam response and with the point sources
subtracted) is 0.67$\pm$0.03 Jy.
This value seems to be significantly lower than
the flux density of 1.4$\pm$0.1 Jy determined by
Thomasson \& Davies (1970) at approximately the same frequency,
but (see below) we are missing flux density in our image. 
The total flux in H$\beta$ for the Helix is
F(H$\beta$) = 3.37 $\times$ 10$^{-10}$ ergs cm$^{-2}$ s$^{-1}$
(O'Dell 1998). Assuming optically-thin emission at
1.4 GHz and in H$\beta$, and an electron temperature of
10$^4$ K, the radio and optical fluxes are related by

\begin{equation}
\Biggl[\frac{S(1.4~GHz)}{Jy}\Biggr]=3.2 \times 10^9~\Biggl[\frac{F(H \beta)}{ergs~cm^{-2}~s^{-1}}\Biggr].
\end{equation}

We then expect, from the H$\beta$ flux, S(1.4~GHz)$\simeq$1.1 Jy.
This estimate also gives a larger value than our
measurement. We believe that the explanation resides in that,
given the large extension of the Helix, we may be losing flux
density as a result of our lack of data taken with very short
baseline separations. Since smooth, extended structures
are detected only with very short spacings, this undersampling
makes interferometers ``blind'' to such structures.
Indeed, for our observational parameters we expect to start
being insensitive to structures larger that 15$'$, which is
the approximate extent of the classic bright ionized nebula. We most probably
do not detect very extended, faint structures, such as those
reported in H$\alpha$ by Speck et al. (2002).

Our continuum image shows that the inner part of the nebula
also emits free-free radiation. In Figure 2 we show a slice
made with PA = 30$^\circ$,
that goes through the position of the central star and
the brightest parts of the ionized shell. 
This slice is very similar in shape to the H$\beta$ slice shown by
O'Dell (1998). As this and other authors have
recently discussed, we find that the central
region is not an empty cavity, but does contain ionized gas.

\subsection{H I Observations}

To study the H I emission we produced natural-weight images 
with the continuum subtracted.
The resulting images have a typical rms of about 1 mJy beam$^{-1}$.
We detect H I emission in a region that seems to surround 
the ionized parts of the nebula. In Figure 3 we show a contour image 
of the H I emission, integrated in LSR radial velocity 
from $-$3.4 to $-$54.9 km s$^{-1}$, the 
interval over which line emission is clearly detected.
The synthesized beam of this image has dimensions of
43$\rlap.{''}6$$\times$41$\rlap.{''}8$; PA=
54$\rlap.^\circ$0).
there is probably additional H I emission at LSR velocities below $-$3.4
km s$^{-1}$,
but contamination from extended H I along the line of sight
makes the $-$3.4
to $+$8.2 km s$^{-1}$ range very uncertain. 
This contour image is overlapped on
the 1.4 GHz continuum free-free emission (greyscale).
We conclude that the H I delineates the outer parts of the bright 
H~II ring that
characterizes NGC~7293. In particular, we do not detect H I in the
region between 100$''$ and 200$''$ from the central star, where 
most of the cometary knots 
are located (O'Dell \& Handron 1996; Meaburn et al. 1998;
L\'opez-Mart\'\i n et al. 2001).
%We can then rule out as an origin for this atomic gas
%photoevaporation from the cometary knots
%in this inner region, since then
%a closer physical association would be expected.
The observed H~I is most likely the outer atomic envelope of the nebula,
most probably photodissociated by the central star.
To have a more accurate comparison between the neutral atomic hydrogen 
and the ionized gas, we made a
a comparison of our H~I image with the STScI digitized
red image of the Second Palomar Observatory Sky Survey (POSS-II).
We assumed that this red image is dominated by line emission
from ionized species
and that approximately traces the ionized gas.
%Our result also supports the conclusion that the nebula is
%ionization bound at least in the plane of the disk.
The astrometry of the image was made with the task XTRAN
of AIPS.
The comparison (see Figure 4) indicates that, although there is no
detectable H~I emission in the bright, inner parts
of the ionized nebula, there is faint optical emission
that seems to coexist with the H~I.

We note here that the presence of continuum emission is not expected to affect
our line results. At a given velocity $v$, the brightness
temperature of the line emission, $T_L$, is given by

\begin{equation}
T_L(v) = (T_{ex} - T_{bg}) (1 - e^{-\tau(v)}),
\end{equation}
where $T_{ex}$ is the excitation temperature of the
transition, $T_{bg}$ is the background brightness temperature, and 
$\tau(v)$ is the opacity at the velocity $v$.
The largest beam-averaged brightness temperature in the 1.4 GHz image 
is $\sim$3 K, and we can safely assume $T_{ex} >> T_{bg}$, since
the excitation temperature of H I in similar environments is
$\geq$60 K. 

\subsection{H~I in absorption}

We also searched for H~I in absorption toward the continuum
sources in the field. Absorption features at
v$_{LSR} \simeq$ 0 km s$^{-1}$ were detected toward
several of these sources. These features are
most probably due to local H~I in the line of sight.
Only toward the most intense source
(with a flux density of $\sim$140 mJy),
located at the SE of the Helix
(at $\alpha(2000) = 22^h~ 29^m~ 
57\rlap.{^s}9;~ \delta(2000) = -20^{\circ}~54{'}28\rlap.{''}0$,
see Figure 1) were we able to detect a statistically significant
feature at velocities away from 0 km s$^{-1}$.
This feature, at a velocity of $\sim -$11.1 km s$^{-1}$,
is shown in Figure 5. The opacity implied is about 0.03,
and it is unclear if this weak absorption is associated
with the Helix or with a line of sight component.

\section{Discussion}

\subsection{The Mass in H I}

In Figure 6 we show a spatially integrated H I spectrum.
Since the largest beam-averaged brightness temperature in the
line is $\sim$6 K (for a feature at $v_{LSR}$ = $-$15.0 km s$^{-1}$,
see below) and
since $T_{ex} \geq$60 K, we can safely conclude from the equation above
that the H~I emission is optically-thin. 

Under optically-thin conditions, the mass in H I is given by

\begin{equation}
\Biggl[\frac{M(H I)}{M_\odot}\Biggr] =~ 0.23~ \Biggl[\frac{D}{kpc}\Biggr]^2~
\Biggl[\frac{\int S_v dv}{Jy~km~s^{-1}}\Biggr],
\end{equation} 
where $D$ is the distance to the object and $S_v$ is the flux density
at velocity $v$. From the spectrum in Figure 6 
we obtain $\int S_v dv$ = 7.9$\pm$0.8 $Jy~km~s^{-1}$ and thus
$M(H I)$ = 0.07$\pm$0.01 $M_\odot$.
The mass in ionized gas is estimated to be $\sim$0.36 $M_\odot$
by Young et al. (1999), following
Gathier (1987) and assuming a filling factor
of 0.75 for the ionized gas.
This filling factor is the average value found by Gathier (1987) from
a statistical study of PNe.
However, such a filling factor is most probably too large for
the Helix (thus overestimating the mass in ionized gas).
Indeed, a more realistic value for this mass is that given by
Boffi \& Stanghellini (1994). These authors use
electron densities derived from forbidden lines 
and H$\beta$ flux densities to derive an ionized mass of $\sim$0.074 $M_\odot$
and a filling factor of $\sim$0.005.
Similar small values of the filling factor have been found
for a few PNe (Boffi \& Stanghellini 1994) and H~II regions
(Kantharia, Anantharamaiah, \& Goss 1998).
The mass in the molecular component is estimated to be
$\sim$0.025 $M_\odot$ (Young et al. 1999).
We then conclude that the atomic hydrogen mass is a few
times the mass in molecular hydrogen and similar to that
in ionized hydrogen. This result confirms
the suggestion of Young et al. (1997), derived from observations
of the 609 $\mu$m 
%ground-state fine-structure 
line of neutral carbon toward
the west limb of the nebula, that a significant atomic envelope
was expected. 

%Another PN for which the ionized, atomic, and molecular masses
%are determined is NGC~6302, for which values of
%0.2, 0.05, and 0.5 $M_\odot$, are derived respectively
%%(G\'omez et al. 1989). 

In Figure 7 we show a panel with the H~I emission at different velocities.
For this panel, the data were further smoothed to a velocity resolution
of 3.9 km~s$^{-1}$.
As it is the case in the CO emission (Young et al. 1999), the H~I emission
exhibits complex kinematics and is rather clumpy.
Unfortunately, the modest signal-to-noise ratio of the H~I emission
precludes a detailed comparison with the CO data of
Young et al. (1999).
The H~I spectrum shown in Figure 6 is characterized 
by a wide component ($\Delta v \simeq$ 40 km~s$^{-1}$)
with a narrow ($\Delta v \simeq$ 3 km~s$^{-1}$)
spike at $v_{LSR} \simeq$ $-$15 km~s$^{-1}$.
This narrow component is located to the NW of the nebula, as can be seen
in Figure 7.

\subsection{Comparison of the H~I with molecular data}

The morphology of the H~I emission shown in Figure 3 is, in general terms,
similar to that observed in CO (see Figure 1 of Young et al. 1999)
and in the 2.122 $\mu$m $v = 1 \rightarrow 0$ s(1)
line of H$_2$ (see Figure 6 of Speck et al. 2002, reproduced
in our Figure 8). 
In particular, once we account for the different angular resolutions
of the images, there is remarkable morphological similarity
between the H~I and the H$_2$ in the outer parts of the
nebula. This can be seen
in Figure 8, where we show an overlap of the H~I emission on the
H$_2$ emission image of Speck et al. (2002).

However, both in
CO and H$_2$ there is significant emission from the inner regions
of the nebula, where the cometary globules exist.
This emission is absent in H~I.
This result could be due in part to the modest sensitivity of
our H~I image, but there are features that are strong in
CO and H$_2$ but remain undetected in H~I. The most evident
case is a bright CO and H$_2$ component about 200$''$ to the
SE of the central star, that has no H~I counterpart.
A possible explanation for the lack of H~I in association with
the cometary globules (that are clearly associated with
both ionized and molecular gas) is that in these objects, as a whole,
there is much less atomic hydrogen than in the outer envelope.
Estimates show that, indeed, more H~I emission is expected from the
outer envelope than from the cometary globules.
Photodissociated regions are characterized by a layer
of atomic hydrogen that extends to a depth of $A_V \sim 1$, that is,
a hydrogen column density of $N_H \sim 2 \times 10^{21}~cm^{-2}$
(Hollenbach \& Tielens 1997).
Then, the H~I mass in a region around a star will be approximately
given by
\begin{equation}
\Biggl[\frac{M(H I)}{M_\odot}\Biggr] =~ 0.16~
\Biggl[\frac{\Omega}{str}\Biggr]~
\Biggl[\frac{r}{0.1~pc}\Biggr]^2,
\end{equation} 
where $\Omega$ is the solid angle subtended by the
H~I zone with respect to the star and $r$ is the characteristic
radius of the H~I zone with respect to the star.

For the outer H~I region we have $r \simeq$ 0.4 pc.
From the geometric model of Meaburn et al. (1998),
we estimate that the disk occupies a solid angle
of order 1 with respect to the star. However, the medium
seems to be clumpy and incomplete (see Figures 3
and 7) and we crudely assume that
$\sim$10\% of this solid angle is actually covered by gas that intercepts
the ultraviolet radiation from the star. We then assume $\Omega \sim$ 0.1.
We then crudely estimate $M(H~I)_{outer} \sim$0.3 $M_\odot$, which is
larger but of the same order of magnitude that the measured
value of 0.07 $M_\odot$.

Now, for the location of the cometary globules we take
$r \simeq$ 0.15 pc. The globules have a characteristic
diameter of 0.001 pc (Dyson et al. 1989)
and it is estimated that there are about 3500 of them
in the Helix (O'Dell \& Handron 1996). 
From these numbers, we estimate
that $\Omega \sim$ 0.01.
We then conclude that $M(H~I)_{inner} \sim$0.03 $M_\odot$,
about ten times less than the value estimated
for the outer region and 
below our detectability limit. 
More sensitive H~I observations would be required to
establish how much atomic hydrogen is present in the
inner regions of the nebula.

To provide a more quantitative comparison between the
spatial distribution of the ionized, atomic, and molecular
hydrogen, we made (see Figure 9) a plot of the average
intensity of the tracers as a function of distance from
the central star. These average intensities are
taken in concentric rings, each with a width of 50$''$, using
the task IRING of AIPS.
A positive bias was subtracted from the red and H$_2$ images,
assuming that this subtraction also removes, on the
average, the contribution
of the stars in the fields.
All three average intensities were normalized so that
their highest value equaled 1.
As can be seen in Figure 9, the ionized gas emission is quite
bright in the central regions, rising in intensity to reach a
peak at $\sim$250-300$''$ from the central star.
After this distance, the intensity has a
gradual decline as a function of distance from the center.
The plot from the red POSS-II image
is consistent with the free-free cut shown in our
Figure 2 and with the H$\beta$ slice shown by
O'Dell (1998). 
Both H~I and H$_2$ are not detected at the center of the
nebula. At more distant radii,
the H$_2$ has a first peak at 
$\sim$200-250$''$ from the central star
and then both H~I and H$_2$ peak at a radius of $\sim$350$''$.
The H$_2$ mosaic becomes unreliable beyond $\sim$500$''$
because of incompleteness and
shorter integration times in that outer region
and its average intensity is plotted only to a radius of $\sim$500$''$.
The average intensity of all three species drops quickly
beyond $\sim$400-500$''$.

The Helix nebula shows, at the same time, evidence of large scale
stratification (i. e. the inner part of the nebula seems to contain
only ionized gas), while at the same time showing coexistence of
several species (in the outer parts ionized, atomic, and molecular
hydrogen are well correlated). We suggest that this situation
could be understood in terms of an inhomogeneous envelope, originally formed
by a large number of neutral globules. Since the destruction time of identical
globules as a result of photoevaporation is roughly inversely proportional
to their distance from the central star (St\"orzer \& Hollenbach 1999),
the globules in the inner regions will soon become photoionized
and form a relatively smooth continuum. 
At intermediate distances from the central star, 
there is some evidence that
the original prominent inhomogenities will survive, leading
to a coexistence of neutral and ionized species in well defined
globules.
Finally, in the outer parts of the nebula most of the globules
could still be present, resulting in a somewhat smoother region where 
ionized, atomic, and molecular hydrogen contributions are present.
The remarkable spatial coincidence of the H~I with the CO and
H$_2$ in the outer ring, where significant ionization
is also still present suggests that most of the observed atomic and
molecular emissions arise in PDRs at the surfaces of knots and
inhomogenities and not because of the presence of a global ionization
front. 
Our model does not account for the fact that
there is also H$_2$ in an inner ring that seems to be separated
from the outer molecular ring. Between these
rings there is a gap in the molecular emission.
It is possible that mass-loss variations during the AGB phase 
(Vassiliadis \& Wood 1993; Steffen, Szczerba, \& Schoenberner 1998;
Speck, Meixner, \& Knapp 2000)
could lead to these concentric structures.

%We conclude that the Helix has at least two zones with neutral gas: one
%that is internal to the ionized gas, 
%related to the cometary globules, and detectable in
%CO and H$_2$, and another that is external to the ionized
%region and detectable in CO, H$_2$, and H~I.
%This difference may reflect the different physical
%conditions of the two zones.
%The two zones are actually evident in the H$_2$ image
%of Speck et al. (2002, their Figure 6), with an inner, very clumpy
%ring associated with the cometary globules, and a smoother, outer
%ring of similar morphology to that of the H~I emission.
%Finally, the presence of these two zones may explain why
%some observers conclude that the molecular gas is
%internal to the ionized gas (Speck et al. 2002), while others
%propose that it is external (Young et al. 1999). There seems
%to be molecular gas in both zones,
%although its nature and origin is probably different. 
%H~I is detected only in the external zone.

\section{Conclusions}

Our main conclusions can be summarized as follows.

1. We report the detection of 21 cm H~I emission from the
Helix Nebula. The mass in atomic hydrogen is estimated to be
$M(H I)$ = 0.07$\pm$0.01 $M_\odot$, about three times larger than
the mass in molecular gas and comparable to the mass
in ionized hydrogen.
The H~I emission covers a range of $\sim$50 km s$^{-1}$ in radial
velocity.

2. The H~I emission seems to delineate the outer parts of the bright 
H~II ring of ionized gas that characterizes
this object. While CO and H$_2$ are found both in this
outer region as well as in a region internal to the ionized gas
and related to the cometary globules, H~I is detected
only in the external region.
We discuss a possible explanation for this difference.
We believe that our observations are consistent with an
inhomogeneous envelope, where the different hydrogen species 
can approximately coexist in some regions
and show in these regions
the same large-scale morphology.
However, there are also significant differences, that we 
tentatively attribute to
variations with distance from the central star, both in the nature of the 
ensemble of globules and in their exposure to ionizing
and photodissociating radiation.

\acknowledgments

We thank Jim Condon, Thierry Forveille, Will Henney, Patrick Huggins,
and Alberto L\'opez  for their valuable comments.
Angela Speck kindly provided us with her H$_2$ image of NGC~7293
and refereed the paper.
LFR is grateful to the support of CONACyT, M\'exico and DGAPA, UNAM.
The Digitized Sky Surveys were produced at the Space Telescope Science 
Institute. The Second Palomar Observatory Sky Survey (POSS-II) was 
made by the California Institute of Technology with funds from 
the National Science Foundation, the National Geographic
Society, the Sloan Foundation, the Samuel Oschin Foundation, and 
the Eastman Kodak Corporation.

\clearpage


\begin{figure}
\plotone{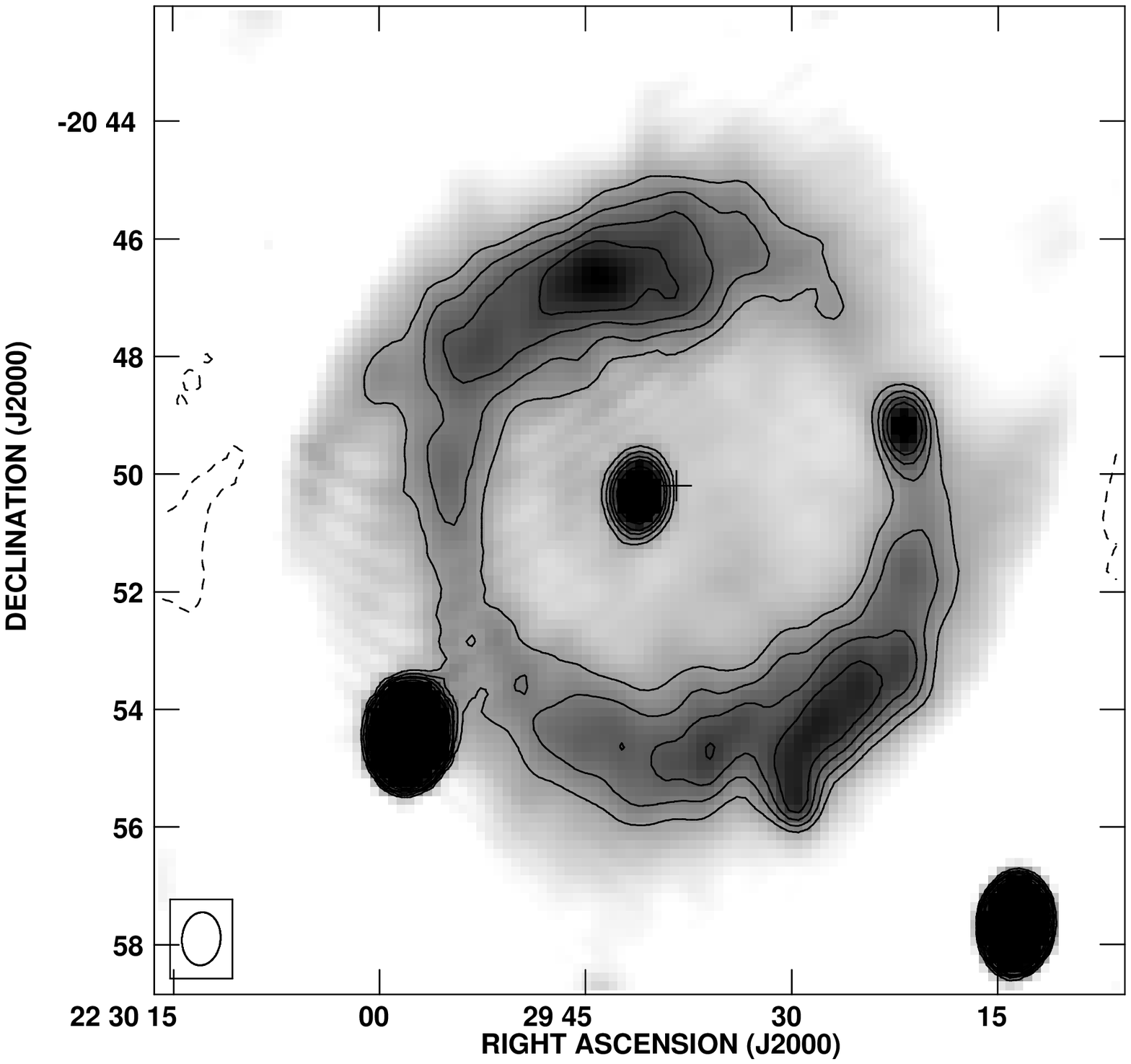}
\caption{VLA continuum image of the Helix Nebula at 1.4 GHz, made with natural
weighting.
Contours are $-$3, 3, 4, 5, 6, 8, 10, 20, 40, and 80 times
1.1 mJy beam$^{-1}$, the rms noise of the image.
The greyscale ranges from 0.0 to 8.5 mJy beam$^{-1}$.
The bright compact objects in the image are most probably
background extragalactic sources.
In particular, we note that the relatively bright radio source
($\sim$25 mJy) near the center of the nebula is not coincident with the 
stellar nucleus.
The position of this
stellar optical nucleus, $\alpha(2000) = 22^h~ 29^m~ 
38\rlap.{^s}4;~ \delta(2000) = -20^{\circ}~50{'}12\rlap.{''}0$,
was taken from Condon \& Kaplan (1998)
and is marked with a cross.
The beam (54$\rlap.{''}2$$\times$39$\rlap.{''}3$; PA=
-5$\rlap.^\circ$0) is shown in the bottom left corner.
\label{fig1}}
\end{figure}

\clearpage 

\begin{figure}
\plotone{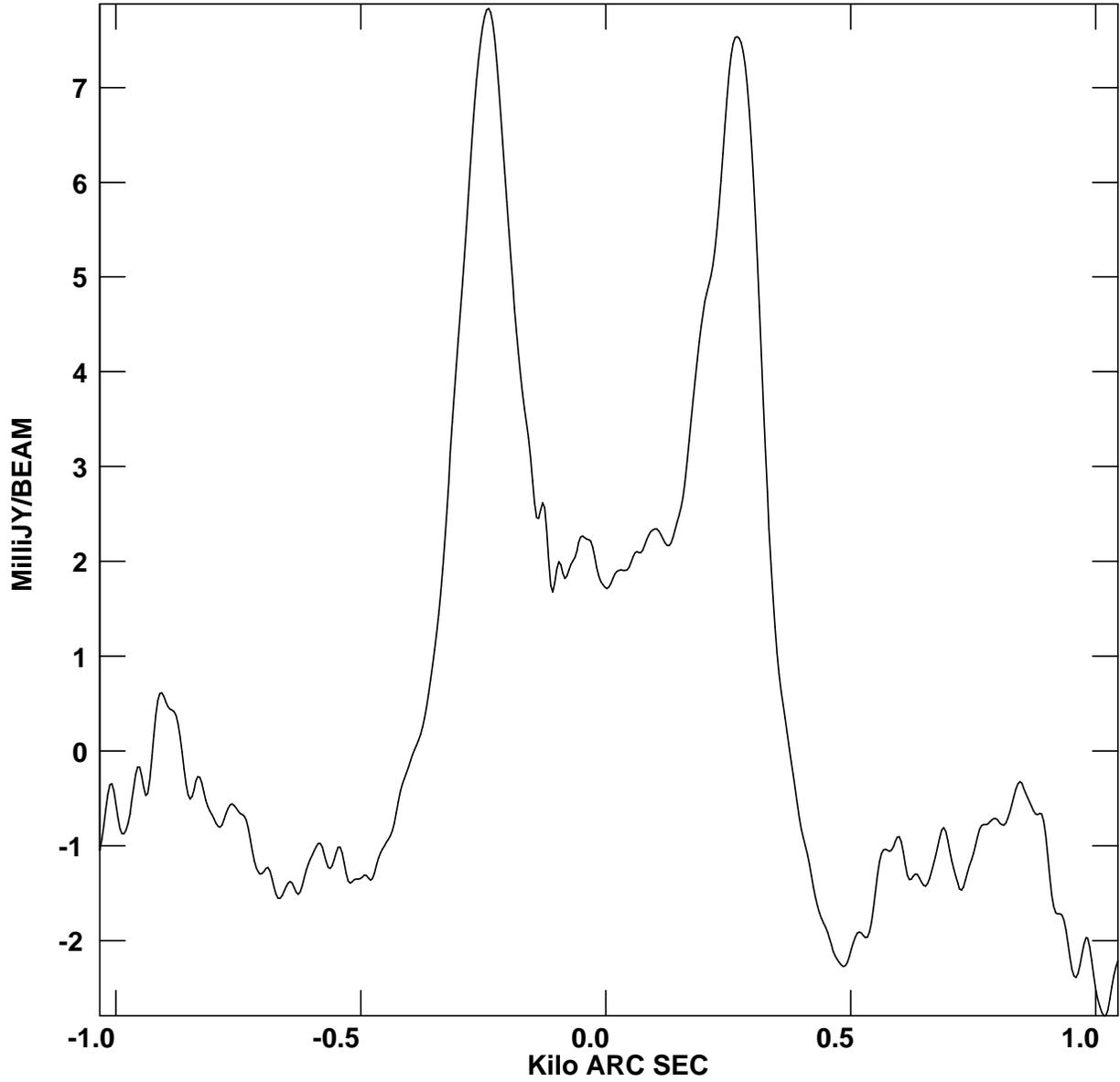}
\caption{Slice made from the continuum
image shown in Figure 1. The slice goes through the
position of the central star at PA = 30$^\circ$.  
Note the presence of negative ``bowls'' to the side
of the source, that as discussed in the text are
indicative of missing flux density.
\label{fig2}}
\end{figure}

\clearpage

\begin{figure}
\plotone{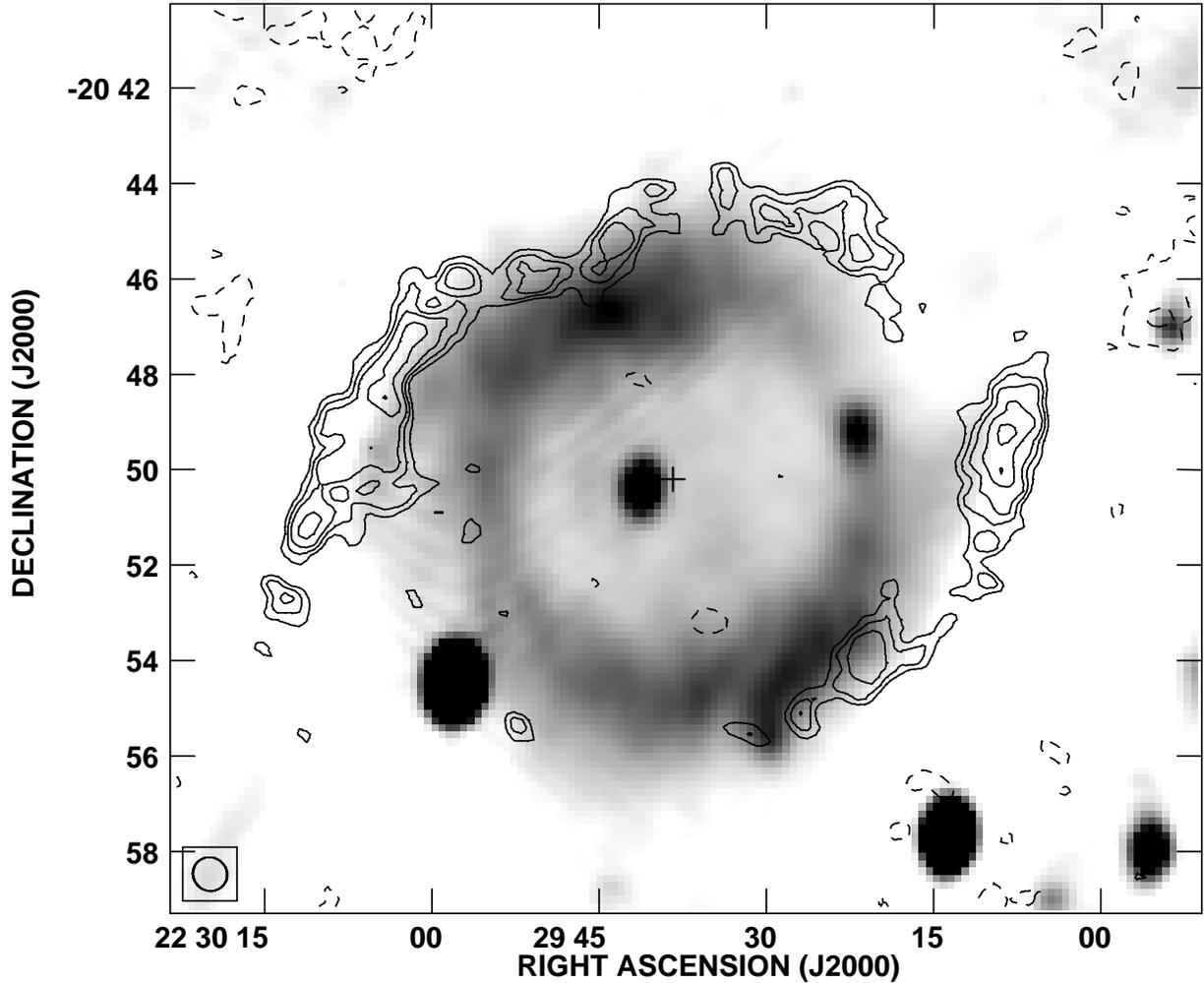}
\caption{Contour image of the H I emission from
the Helix Nebula at 1.4 GHz, integrated from
$-$3.4 to $-$54.9 km s$^{-1}$. The image was made with natural
weighting.
Contours are $-$4, 4, 5, 6, 8, and 10 times
13.4 mJy beam$^{-1}$ km s$^{-1}$, the rms noise of the integrated image.
The continuum emission is shown
in greyscale, with a range from 0.0 to 8.5 mJy beam$^{-1}$.
The H~I beam (43$\rlap.{''}6$$\times$41$\rlap.{''}8$; PA=
54$\rlap.^\circ$0) is shown in the bottom left corner.
The cross marks the position of the
stellar nucleus of the planetary nebula.
\label{fig3}}
\end{figure}

\clearpage

\begin{figure}
\plotone{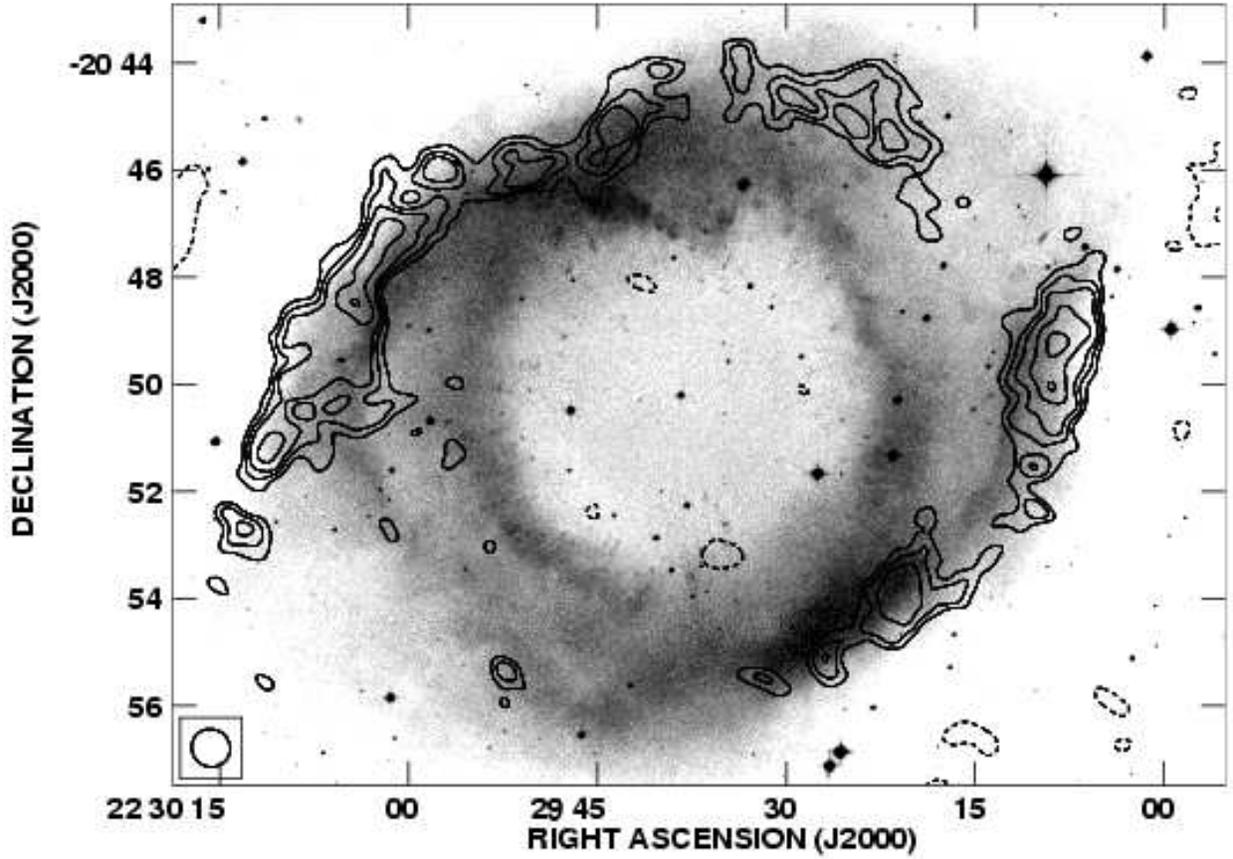}
\caption{Contour image of the H I emission from
the Helix Nebula, as in Figure 3,
superposed on the red image of the POSS-II
(greyscale).
The greyscale range goes from 0.64 to 0.89 of the peak value
of the image.
The H~I emission appears associated with the outer parts of the ionized
nebula.
\label{fig4}}
\end{figure}

\clearpage

\begin{figure}
\plotone{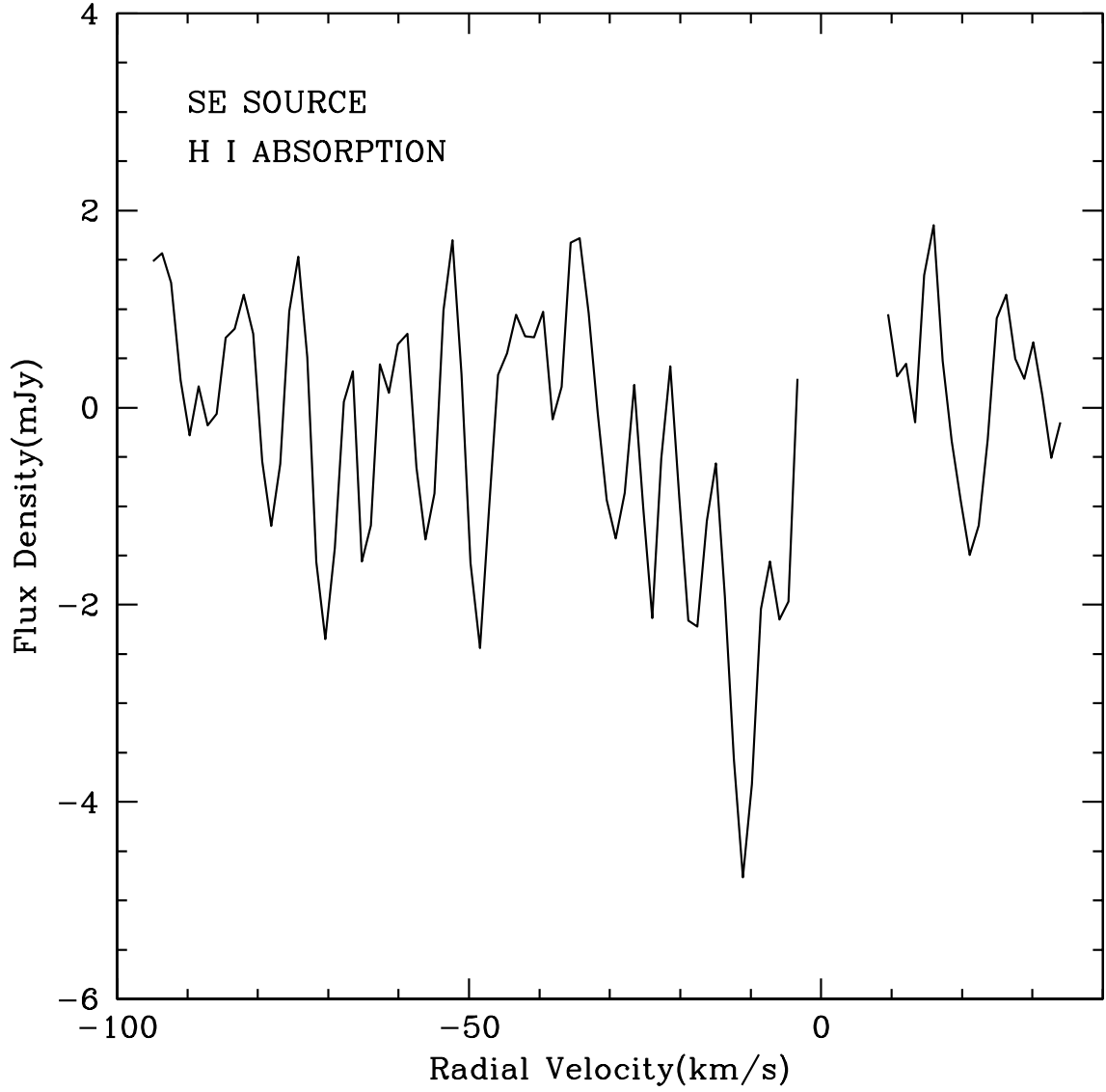}
\caption{H~I absorption spectrum toward the bright
(with a flux density of 140 mJy) 
radio continuum source at the SE of the Helix,
at $\alpha(2000) = 22^h~ 29^m~ 
57\rlap.{^s}9;~ \delta(2000) = -20^{\circ}~54{'}28\rlap.{''}0$
(see Figure 1).
\label{fig5}}
\end{figure}

\clearpage

\begin{figure}
\plotone{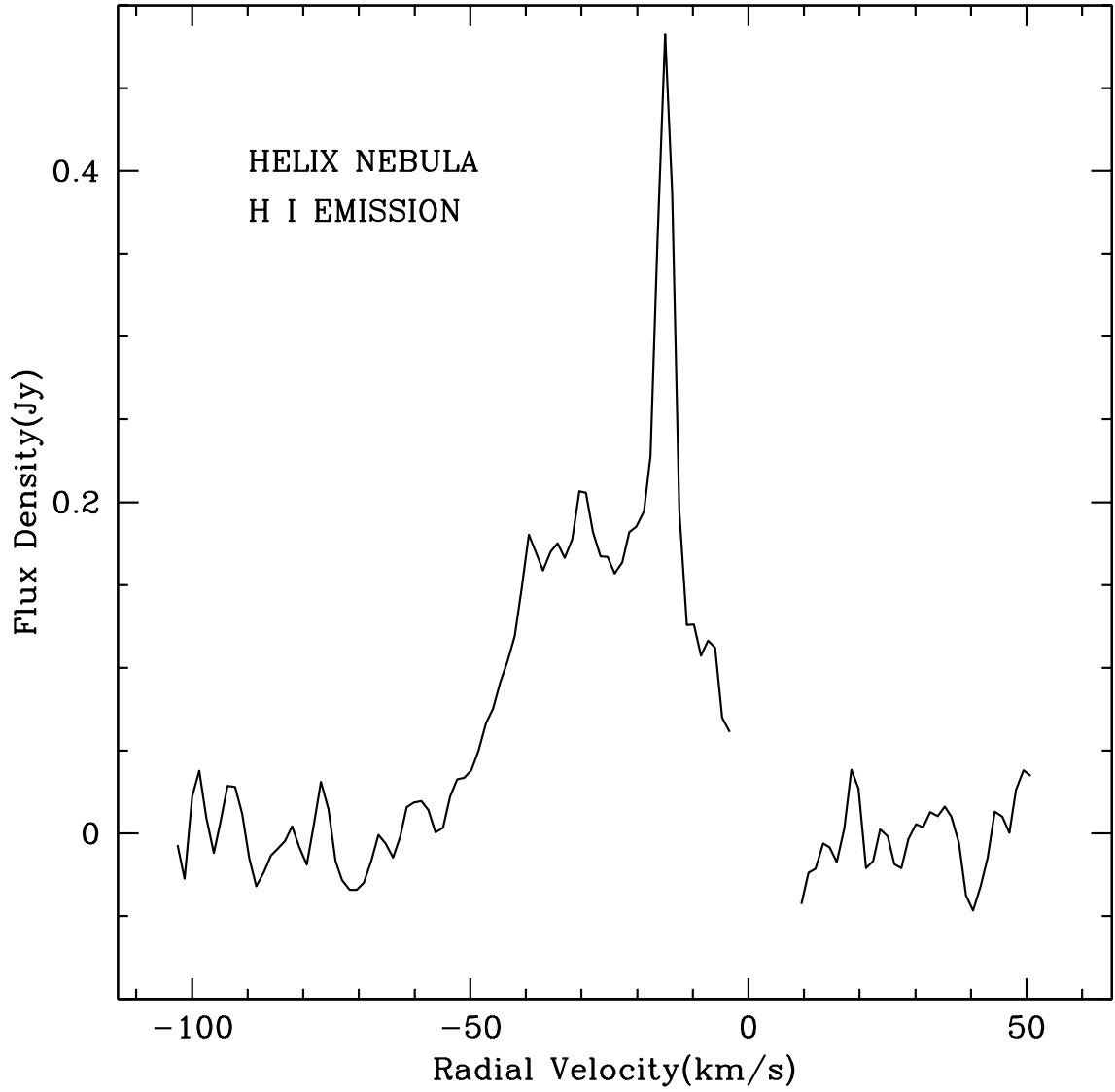}
\caption{
Spatially-integrated H~I spectrum toward the Helix Nebula, corrected for
primary beam response. A linear baseline has been subtracted.
The region between $-$3.4 to $=+$8.2
km s$^{-1}$ is not shown since
it is contaminated with emission from extended line-of-sight H I.
\label{fig6}}
\end{figure}

\clearpage

\begin{figure}
\plotone{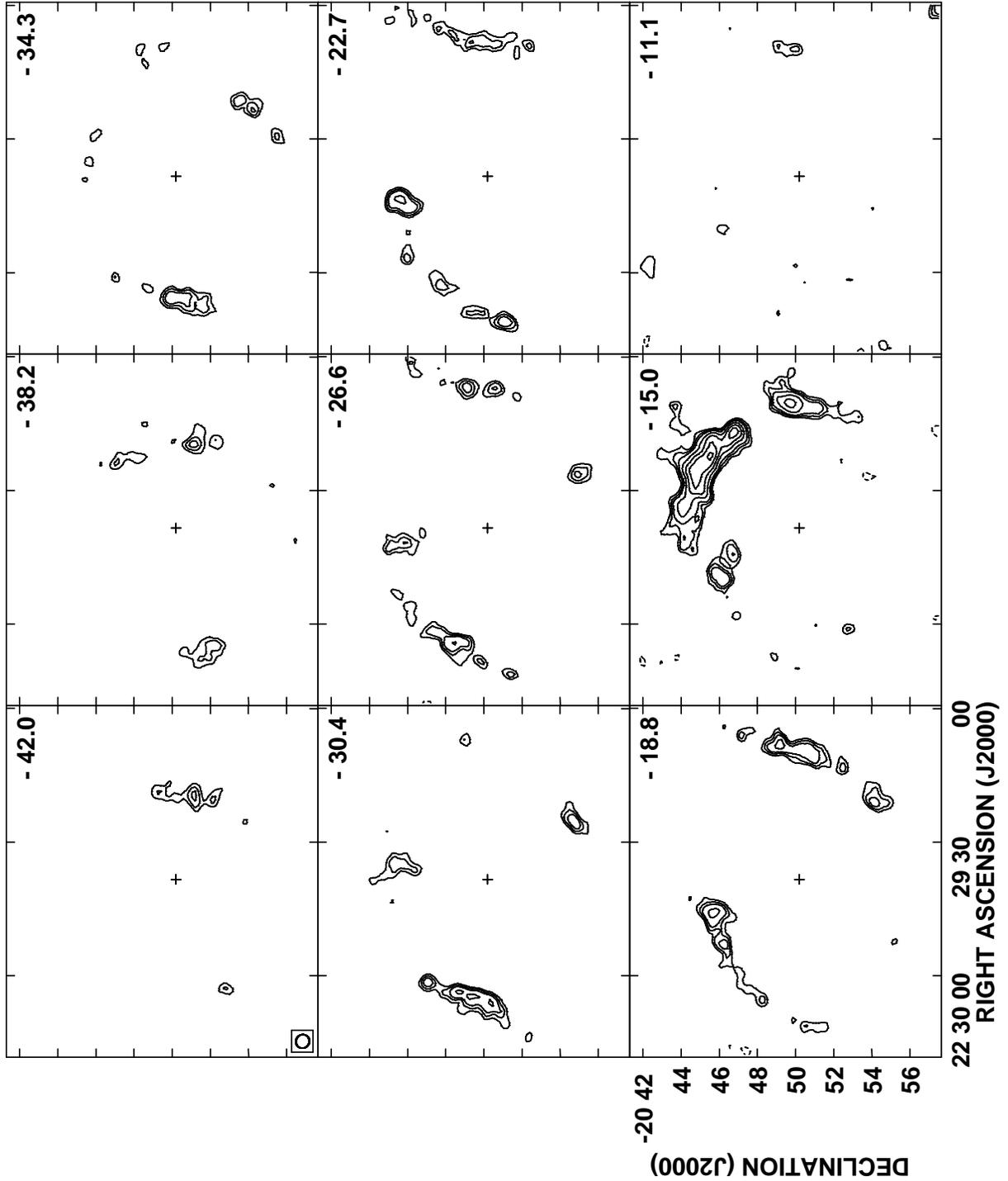}
\caption{Panel with contour images of the H~I emission for the
LSR velocities indicated in the top right corner of each image. 
The velocity resolution of these images is 3.9 km s$^{-1}$.
Contours are $-$4, 4, 5, 6, 8, 10, 12, and 14 times
1.0 mJy beam$^{-1}$, the average rms of the images.
The beam (43$\rlap.{''}6$$\times$41$\rlap.{''}8$; PA=
54$\rlap.^\circ$0) is shown in the top left image.
The cross marks the position of the
stellar nucleus of the planetary nebula.
\label{fig7}}
\end{figure}

\clearpage

\begin{figure}
\plotone{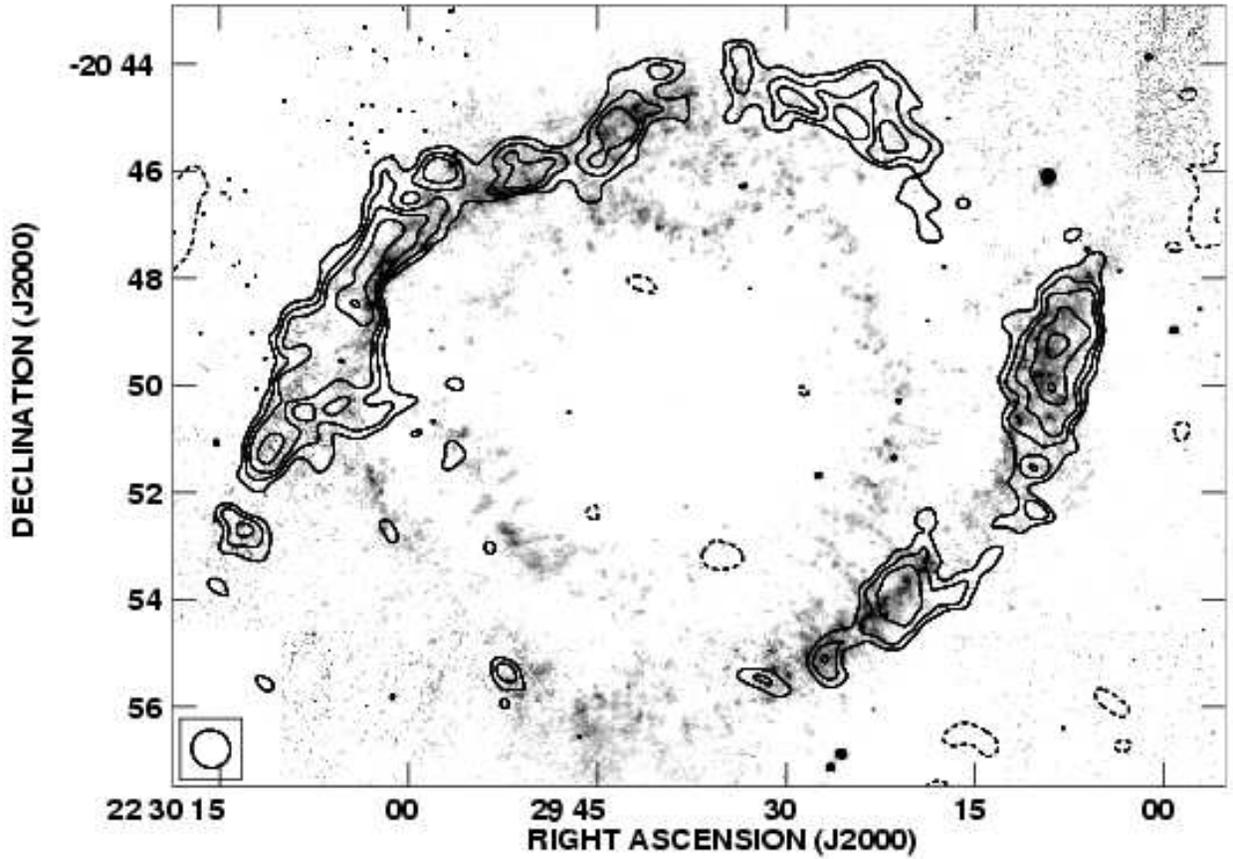}
\caption{Contour image of the H I emission from
the Helix Nebula, as in Figure 3,
superposed on the H$_2$ image of Speck et al. (2002),
shown in greyscale.
The greyscale range goes from $1.0 \times 10^{-4}$
to $3.0 \times 10^{-4}$ ergs s$^{-1}$ cm$^{-2}$ sr$^{-1}$.
Note the remarkable morphological similarity between the
H~I and the H$_2$ in the outer parts of the nebula.
\label{fig8}}
\end{figure}

\clearpage

\begin{figure}
\plotone{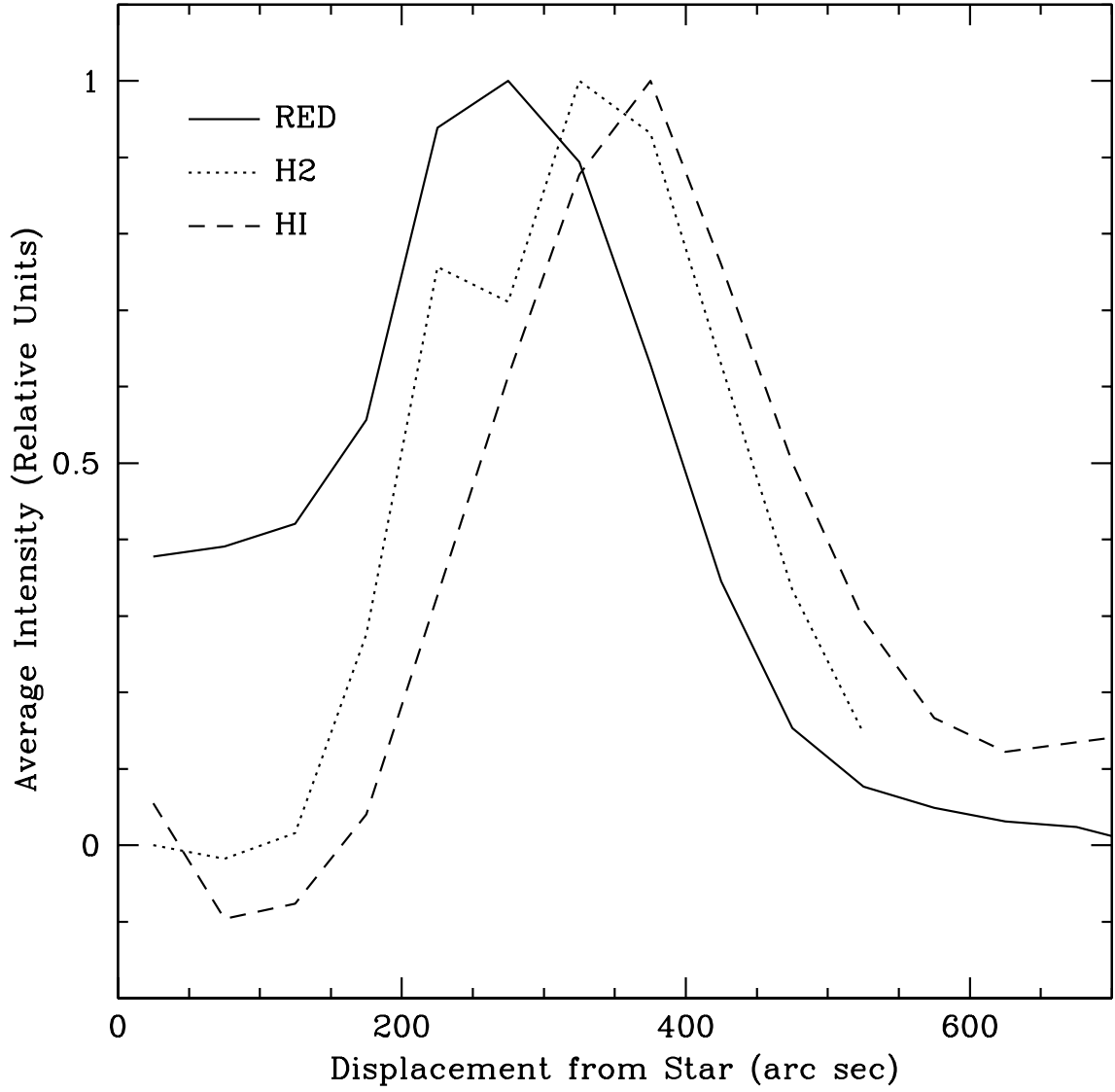}
\caption{Average intensity, in normalized units,
of the tracers of ionized (red image), atomic (H~I),
and molecular (H$_2$) hydrogen in the Helix nebula,
as a function of radius from the central star.
\label{fig9}}
\end{figure}

\end{document}